\documentclass[%
 reprint,
superscriptaddress,
 amsmath,amssymb,
prapplied
]{revtex4-2}
\usepackage{graphicx}
\usepackage{dcolumn}
\usepackage{bm}

\usepackage{siunitx}
\usepackage[shortconst]{physconst}

\usepackage{amsmath}
\usepackage{graphicx}
\usepackage[colorlinks=true, allcolors=blue]{hyperref}

\begin{document}

\def\figureautorefname{Fig.}
\def\tableautorefname{Tab.}
\def\equationautorefname{Eq.}

\title{High-$Q$ superconducting resonators fabricated in an industry-scale semiconductor-fabrication facility}

\author{N. Arlt}
\email{nicolas.arlt@infineon.com}
\author{K. Houska}%
 \email{karina.houska@infineon.com}
\affiliation{%
 Technical University of Munich, TUM School of Natural Sciences\\
 Department of Physics, 85748 Garching, Germany
}%
\affiliation{Infineon Technologies AG, 85579 Neubiberg, Germany}
\affiliation{Walther-Meißner-Institut, Bayerische Akademie der Wissenschaften, 85748 Garching, Germany}
\author{J. Braum\"uller}
\author{M. Kirsch}
\author{G. Metzger-Br\"uckl}
\author{W. Raberg}
\author{T. Stangl}
\affiliation{Infineon Technologies AG, 85579 Neubiberg, Germany}
\author{S. Filipp}
\affiliation{%
Technical University of Munich, TUM School of Natural Sciences\\
Department of Physics, 85748 Garching, Germany
}%
\affiliation{Walther-Meißner-Institut, Bayerische Akademie der Wissenschaften, 85748 Garching, Germany}
\author{J. Banker}
\author{F. Brandl}
\affiliation{Infineon Technologies AG, 85579 Neubiberg, Germany}

\date{\today}

\begin{abstract}
Universal quantum computers promise to solve computational problems that are beyond the capabilities of known classical algorithms. To realize such quantum hardware on a superconducting material platform, a vast number of physical qubits has to be manufactured and integrated at high quality and uniformity on a chip. 
Anticipating the benefits of semiconductor industry processes in terms of process control, uniformity and repeatability, we set out to manufacture superconducting quantum circuits in a semiconductor fabrication facility. In order to set a baseline for the process quality, we report on the fabrication of coplanar waveguide resonators in a \qty{200}{\mm} production line, making use of a two-layer superconducting circuit technology. We demonstrate high material and process quality by cryogenic Q-factor measurements exceeding $10^6$ in the single-photon regime, for microwave resonators made of both Niobium and Tantalum. In addition, we demonstrate the incorporation of superconducting Niobium air bridges in our process, while maintaining the high quality factor of Niobium resonators.
\end{abstract}

\maketitle

\section{Introduction}

The advancement of superconducting quantum computing relies heavily on the ability to scale up quantum circuits without sacrificing individual qubit performance. One central ingredient for this is the fabrication of large-scale circuits at very low parameter spread and reliable parameter targeting \cite{kjaergaard_superconducting_2020}. The semiconductor industry will likely play a crucial role in addressing these challenges, due to its expertise in large-scale, high-precision manufacturing \cite{mohseni_how_2024,van_damme_advanced_2024}. While typical small-scale fabrication methods will likely struggle to meet the stringent demands in terms of uniformity and reproducibility, the high degree of process control on a mass scale established in the semiconductor sector offers a pathway to overcoming these challenges.
 
As a first step towards large-scale production of superconducting quantum circuits, we implement superconducting microwave resonators, an essential component in the superconducting circuit toolbox \cite{blais_cavity_2004,wallraff_strong_2004}, and also an ideal detector for characterizing fabrication quality \cite{martinis_decoherence_2005,mcrae_materials_2020}. We demonstrate the fabrication of coplanar microwave resonators using both Niobium (Nb) and Tantalum (Ta) in a \qty{200}{\mm} semiconductor fabrication facility. In particular, we exclusively rely on productive fabrication tools routinely used for the production of semiconductor devices, in contrast to the usual approach of employing highly specialized tools solely dedicated to the fabrication of superconducting quantum circuits. Our process achieves internal quality factors exceeding $10^6$ in the single-photon regime, indicating high (interface) material and process quality, which demonstrates the general suitability of semiconductor industry tools for the fabrication of superconducting quantum circuits. Furthermore, we incorporate superconducting Nb air bridges \cite{schicke_integrated_2003,wan_fabrication_2024,bruckmoser_niobium_2025}, a widely adopted approach by the community to interconnect circuit elements or suppress unwanted circuit modes by connecting groundplanes, without sacrificing the high Q-factors of the resonators. 

\section{High-Q resonators fabricated in an industrial semiconductor cleanroom}\label{sec1}
\paragraph{Industry scale manufacturing.}
\begin{figure*}[ht]
    \centering
    \includegraphics[scale=1]{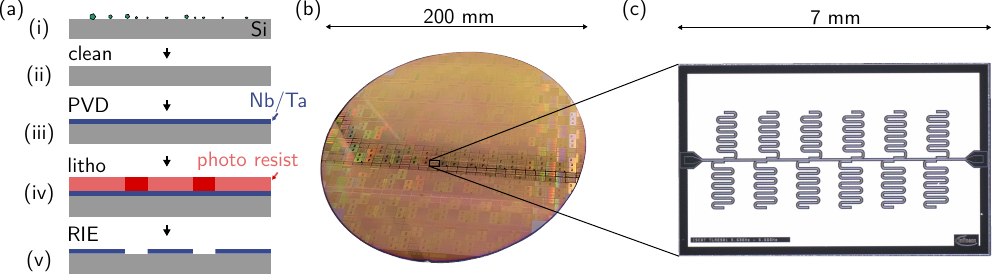}
    \caption{\fontsize{10pt}{12pt}\selectfont\textbf{Fabrication flow of superconducting resonators on $\mathbf{200\,mm}$ wafers} (a) Schematic of fabrication process flow. Contaminations and surface oxides on the blank wafers  (i) are removed by RCA and DHF cleaning processes (ii). Metal films are deposited using physical vapor deposition (PVD), optionally preceded by an in-situ deposition of a thin seed layer (iii). The film is structured using optical lithography (iv) and Cl-based reactive ion etching (v), after which the remaining photoresist is removed using a wet chemical process. (b) Fully processed \qty{200}{\mm} wafers are diced into \qtyproduct[product-units=power]{7x4.3}{\mm} samples and subsequently wire bonded to prepare for cryogenic characterization. (c) Optical micrograph of CPW resonator chip.}  
    \label{fig:fig1}
\end{figure*}
For the fabrication of coplanar microwave resonators, we follow a subtractive, etch-based fabrication flow sketched out in \autoref{fig:fig1}a. As a substrate material, we use \qty{200}{mm} Si (100) wafers of \qtyrange[range-phrase=--,range-units=single]{3}{8}{k\ohm \cm} nominal resistivity, which undergo a semiconductor-industry standard RCA cleaning procedure to remove organic contaminations \cite{kern_evolution_1990}, followed by a dilute HF (DHF) process to remove surface oxides. Cleaned substrates are typically transferred to the deposition tool within $\qty{1}{\hour}$ to prevent excessive re-oxidation. This lies within logistical requirements of the fabrication facility, which is compatible with the logistical boundary conditions of Infineon’s semi-automated fabrication facility. The main metalization layer is subsequently deposited at room temperature using DC sputtering, to a total thickness of \qty{150}{\nano\meter}. Nb films are deposited directly on the silicon substrate, while for Ta, we test three deposition conditions. We either deposit Ta directly on Si or employ in-situ deposition of \qty{5}{\nm} seed layers of either Nb or TaN, to enhance growth of $\alpha$-phase Ta \cite{face_nucleation_1987,gladczuk_sputter_2005}. We later refer to these films as (bare) Ta, Nb-Ta and TaN-Ta, respectively. Multiple targets are available in the tool, so that seed layers are deposited without vacuum break. We employ optical lithography using an i-Line stepper with nominal resolution of feature sizes down to \qty{350}{\nm} on previously applied photoresist. After exposure and resist development, we transfer the resist pattern to the film in a Cl-based reactive ion etch process (RIE), where we use similar process parameters for both Nb and Ta films. Avoiding the use of oxygen plasma, photoresist is removed using a commercially available chemical resist stripper, minimizing surface oxidation. Finally, wafers are diced into chips with area \qtyproduct[product-units=power]{7x4.3}{\mm} (see \autoref{fig:fig1}b,c) and wire bonded into a microwave package for cryogenic characterization.\\
\paragraph{Performance benchmarking and TLS model.}
We quantify the performance of metal films and interface dielectrics by the internal quality factor $Q_\mathrm{i}$ of microwave resonators at cryogenic conditions. 
We use a chip design comprising twelve $\lambda/4$ coplanar waveguide (CPW) resonators of various lengths coupled to a common feedline in the notch configuration \cite{probst_efficient_2015}, as depicted in \autoref{fig:fig1}c. CPW structures are designed to span a frequency range of \qtyrange[range-phrase=--,range-units=single]{5.7}{6.8}{\giga\Hz}, with a center conductor width and gap of \qty{10}{\um} and \qty{6}{\um}, respectively, to maintain a characteristic impedance of approximately \qty{50}{\ohm}. The designed coupling quality factor is $Q_\mathrm{c}\approx 10^5$. We mount the samples at the mixing chamber stage of a BlueFors LD dilution refrigerator that is cooled to below \qty{20}{\milli\kelvin} and measure $S_{21}$ in dependence of applied input powers, using a vector network analyzer (Keysight P5003B). Employing a circle fit routine \cite{probst_efficient_2015} on the complex transmission response around the fundamental resonance frequency, we infer $Q_\mathrm{i}$ for each measured resonator structure in dependence of its average photon population $ n$, which we compute based on signal input power $P_\mathrm{in}$ as $n=\frac{2}{\hbar\omega_\mathrm{r}^2}\frac{Q_\mathrm{l}^2}{Q_\mathrm{c}}P_\mathrm{in}$ \cite{bruno_reducing_2015,mcrae_materials_2020,urade_microwave_2024}. Here, $Q_\mathrm{l}$ and $Q_\mathrm{c}$ are the loaded and coupling quality factors, respectively, as obtained from the fitting routine. $\hbar$ is the reduced Planck constant and $\omega_\mathrm{r}$ is the resonator frequency. $P_\mathrm{in}$ is estimated based on wired attenuators in the input line, excluding cable losses, so that \autoref{eq:TLS_loss} provides an upper limit to the number of photons populating the resonators.\\
\begin{figure*}[ht] 
    \centering
    \includegraphics[scale=1]{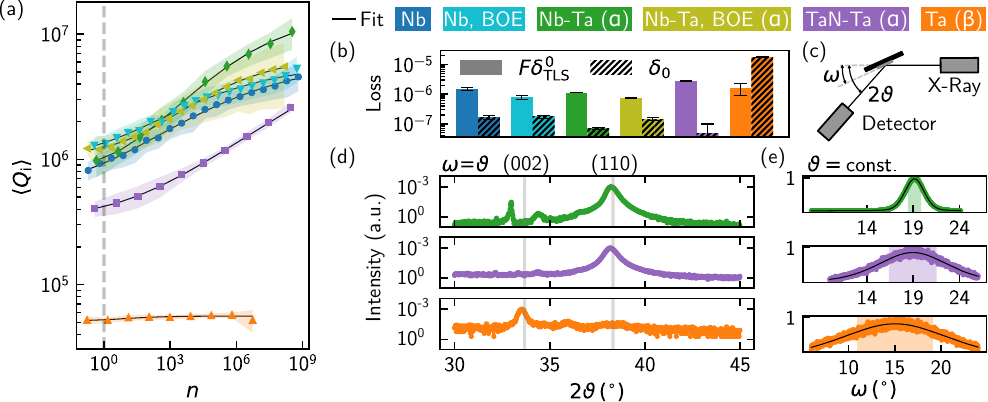}
    \caption{\fontsize{10pt}{12pt}\selectfont\textbf{Characterization of Nb and Ta thin films} (a) Internal quality factor measurements of Nb and Ta CPW resonators. Data points show the average photon-number dependent internal Q-factor $\langle Q_\mathrm{i}\rangle$, with standard deviation indicated by shaded regions. Black solid lines show fit to TLS model (Eq. \eqref{eq:TLS_loss}). Grey dashed line at $n=1$ indicates the single-photon regime. (b) Effective resonator loss contributions as obtained from fitting the power dependence of $\langle Q_\mathrm{i}\rangle$ to Eq. \eqref{eq:TLS_loss}. (c) Schematic of the XRD setup with incident angle $\omega$ and detector angle $\theta$. (d) XRD-analysis of Ta films. Both source and detector are rotated, so that $\omega=\theta$. The bare Ta film without seed layer (orange data) shows a strong peak characteristic to the tetragonal $\mathrm{\beta}$-phase, whereas the presence of both TaN (purple) and Nb (green) seed layers results in a prominent bcc (110) peak, suggestive of the presence of the $\mathrm{\alpha}$-phase. (e) Rocking curves of dominant Ta peaks. The detector angle $\theta $ is fixed at XRD peak, while incident angle $\omega$ is varied. Shaded region visualizes peak width obtained by fit to Pseudo-Voigt profile (black solid lines), indicating highest crystallinity of Ta film on Nb seed layer. Fitted profile widths are \ang{1.4}, \ang{5.2},  and \ang{8.2} from top to bottom, respectively. (d,e) XRD intensity is normalized to maximum value in respective traces.}  
    \label{fig:fig2}
\end{figure*}

Measurements show internal quality factors in the single-photon regime of approximately $10^6$ for the Nb and Nb-Ta films, as shown in \autoref{fig:fig2}a. Data points depict the averaged internal Q-factors $\langle Q_\mathrm{i}\rangle$, where the average is taken over the measured $Q_\mathrm{i}$ of individual resonators, grouped together by their photon population $n$. Shaded regions indicate the corresponding standard deviations. We show measurements of one Nb and three different Ta films, where Nb data is taken from three individual samples. Individual circle fit traces with fitting errors beyond \qty{20}{\percent} were discarded for the analysis. 

All resonators except the Ta film without seed layer show a strong photon-number dependence of $\langle Q_\mathrm{i}\rangle$, as expected for two-level system (TLS) dominated resonators \cite{gao_physics_2008,wang_improving_2009,mcrae_materials_2020,crowley_disentangling_2023,urade_microwave_2024}. Both Nb and Nb-Ta films show the best results, while the TaN-Ta film qualitatively follows a similarly pronounced trend, but at an overall decreased $\langle Q_\mathrm{i}\rangle$. 
In contrast, the bare Ta film deposited directly on silicon shows only a very weak dependence on photon population and shows a reduced $\langle Q_\mathrm{i}\rangle$ by one order of magnitude in comparison to the other films. Low power Q-factors $\langle Q_\mathrm{i}^\mathrm{LP}\rangle$ are summarized in \autoref{t1}. \\ 

To further improve the performance of our best films, we treat samples from both the Nb and the Nb-Ta wafers with an additional \qty{20}{\min} long post-processing buffered oxide etch (BOE), a well-established method to remove detrimental surface oxides from the substrate-air and metal-air interfaces \cite{wisbey_effect_2010,verjauw_investigation_2021,altoe_localization_2022}. BOE treated samples were left at ambient conditions for more than \num{3} weeks, and subsequently characterized in the same way as described above. The results are depicted in \autoref{fig:fig2}a,
indicating a long-term improvement of $\langle Q_\mathrm{i}\rangle$ by more than \qty{50}{\percent} and \qty{20}{\percent}, respectively, resulting in $\langle Q_\mathrm{i}^\mathrm{LP}\rangle>10^6$ (see cyan and olive data in \autoref{fig:fig2}a and b). \\
We next analyze the photon-number dependence of the Q-factors quantitatively. For superconducting resonators, $Q_\mathrm{i}$ is directly related to the effective loss tangent via $1/Q_\mathrm{i}=\tan\delta\approx\delta$. Generally, $\delta$ contains contributions from various dissipation mechanisms, but is usually dominated by TLS-induced resonator loss \cite{martinis_decoherence_2005,mcrae_materials_2020}. While remaining agnostic about the precise microscopic origins, TLS loss models describe generic structural defects in amorphous materials, such as surface oxides or interlayer-interfaces, as bath of short-lived quantum mechanical two-state system with an electrical dipole moment coupling to the electric field of microwave resonators, giving rise to dielectric loss
\cite{muller_towards_2019,anderson_anomalous_1972,phillips_two-level_1987,mcrae_materials_2020}.

\begin{table*}[th!]
\centering
\caption{Low-power quality factors and TLS model fit results of differently fabricated resonators, containing data from all measured devices shown in \autoref{fig:fig2} and \autoref{fig:air bridge}. Fit parameters $F\delta_\mathrm{TLS}^0$ and $\delta_0$ represent TLS loss and power-independent losses, respectively. $n_\mathrm{c}$ is a critical photon number, beyond which TLS saturation reduces the effect of TLS-related losses. $\beta$ is a correction factor accounting for the non-uniform electric field distribution in CPW resonators.}
\label{t1}
\begin{tabular}{c|ccccc}
\noalign{\smallskip}\noalign{\smallskip}
&  $\langle Q_\mathrm{i}^\mathrm{LP}\rangle\,(10^5)$ & $F\delta_\mathrm{TLS}^0\,(10^{-7})$ & $\delta_0\,(10^{-7})$ & $n_\mathrm{c}$ & $\beta$ \\
\hline
\hline
Nb & $8\pm 1.4$ & $15\pm 2.0$ & $1.6\pm 0.25$ & $0.2\pm 0.23$ & $0.29\pm0.03$ \\
after BOE & $13\pm 2.3$ & $8\pm 1.3$ & $1.7\pm 0.3$ & $\num{1}\pm\num{1.4}$ & $\num{.30}\pm\num{.05}$ \\
with air bridge & $10\pm 3$ & $14 \pm 1.4$ & $0.9 \pm 0.12$ & $\num{0.3}\pm\num{0.3}$ & $\num{.37}\pm\num{.024}$ \\
\hline
Nb-Ta & $10\pm 2.6$ & $11.3\pm 0.3$ & $0.69 \pm 0.06$ & $\num{3.9}\pm\num{0.8}$ & $\num{.40}\pm\num{.012}$ \\
after BOE & $12\pm 1.7$ & $7.6\pm 0.4$ & $1.5\pm 0.15$ & $\num{16}\pm\num{7}$ & $\num{.37}\pm\num{.03}$ \\
\hline
TaN-Ta & $4.0\pm 0.6$ & $29.1\pm 1.2$ & $0.5\pm 0.5$ & $\num{11}\pm\num{4}$ & $\num{.24}\pm\num{.012}$ \\
\hline
Ta & $0.52\pm 0.04$ & $16\pm 7$ & $178\pm 3$ & $\num{3}\pm\num{7}$ & $\num{.7}\pm\num{.6}$ \\
\hline

\end{tabular}
\end{table*}
A particular hallmark of TLS-mediated loss is its distinct dependence on the available number of photons $ n$ populating the resonator, which can be modeled as \cite{gao_physics_2008,wang_improving_2009,mcrae_materials_2020,crowley_disentangling_2023,urade_microwave_2024}
\begin{equation}\label{eq:TLS_loss}
    \delta( n)=\frac{F\delta_\mathrm{TLS}^0}{\sqrt{1+\left(\frac{ n}{n_\mathrm{c}}\right)^\beta}}+\delta_0,
\end{equation}
assuming that thermal fluctuations leave the resonator in the ground state, i.e. $\frac{\hbar\omega_\mathrm{r}}{k_\mathrm{B}T}\gg1$. The first term of Eq. \eqref{eq:TLS_loss} represents the effective dielectric loss caused by photons coupling to TLS dipole moments contributing with a participation ratio $F$. Here $\delta_\mathrm{TLS}^0$ represents intrinsic TLS losses that saturate once $ n$ exceeds a critical photon number $n_\mathrm{c}$, marking the onset of TLS saturation. The exponent $\beta$ is a geometry-dependent correction factor accounting for the non-uniform electric field in CPW resonator \cite{wang_improving_2009,urade_microwave_2024}. The second term $\delta_0$ contains power-independent contributions to resonator loss. Implicitly assuming that the power dependence of the Q-factor is dominated by effectively one type of TLS, we fit this model to the ensemble averaged data depicted in \autoref{fig:fig2}a, indicated by black solid lines. The observed trends with photon number are in good agreement with the simple TLS model. Extracted fit parameters are summarized in \autoref{t1} and \autoref{fig:fig2}b. Note that for the bare Ta film, we exclude the two highest power data points, as here the resonator becomes nonlinear, resulting in ill-behaved extraction of $Q_\mathrm{i}$. Also note that $n_\mathrm{c}$ exhibits relatively high fitting errors, but should be rather understood as order-of-magnitude at which TLS saturation effects start to play a role, typically in the single- to few-photon regime. We plot the effective TLS-specific loss $F\delta_\mathrm{TLS}^0$ and the power-independent loss contribution $\delta_0$, as extracted from the fit, in \autoref{fig:fig2}b. Nb, Nb-Ta and TaN-Ta devices are limited by $F\delta_\mathrm{TLS}^0$, with TLS related losses in the TaN-Ta film exceeding those in the untreated Nb and Nb-Ta films by more than a factor of \num{2}. In agreement with the trend observed in \autoref{fig:fig2}a, the bare Ta film's losses are strongly dominated by other loss mechanisms, explaining the large uncertainty in $F\delta_\mathrm{TLS}^0$, as well as in the critical photon number $n_\mathrm{c}$ (see \autoref{t1}). In addition, \autoref{t1} and \autoref{fig:fig2}b reveal a more than \qty{30}{\percent} reduction of TLS losses in BOE-treated samples in comparison with untreated samples. This conforms well with results from similar work on Nb resonators \cite{verjauw_investigation_2021,altoe_localization_2022}. \\

\paragraph{X-ray diffraction (XRD) analysis of Ta films.}
We next investigate material properties of the Ta-based films to understand the performance differences evident in the data. We use a Bruker D8 6-Axis X-ray diffractometer with monochromator, employing $\mathrm{K}\alpha_1$ radiation of a copper X-ray source. XRD spectra are obtained by varying incident and detector angles $\omega$ and $\theta$, so that $\omega=\theta$ (see \autoref{fig:fig2}c). \autoref{fig:fig2}d depicts normalized XRD spectra of each analyzed film. We find that XRD traces of Ta films deposited with seed layer reveal strong peaks at a position close to the (110) peak of the cubic $\alpha$-Ta phase at $2\theta=38.32^\circ$ \cite{dhundhwal_high_2025}. This peak is absent in the Ta film without seed layer, which, in contrast, has a large signal close to the (002) reflex of the tetragonal $\beta$-phase at $2\theta=33.68^\circ$ \cite{dhundhwal_high_2025}, not visible in the other films. We note the presence of weak signals in both films with seed layers at $2\theta\approx34^\circ$ and an additional peak in the Nb-Ta film close to $2\theta\approx33^\circ$, and interpret the data as follows: films with Nb/TaN seed layers are dominated by $\alpha$-Ta, while we attribute the enhanced presence of the $\beta$-phase to the absence of a seed layer. This interpretation conforms with previous findings from \cite{face_nucleation_1987,urade_microwave_2024,crowley_disentangling_2023,dhundhwal_high_2025}. The observed absence of $n$-dependence of Q-factors in the $\beta$-Ta film agrees well with results from \cite{urade_microwave_2024}, where it is hypothesized that dominant photon number-independent losses are caused by the enhanced presence of quasiparticles due to the low critical temperature and poor crystallinity of such films.\\
To examine the crystal quality of the Ta films, we additionally measure the relative distribution of crystal orientations in the polycrystalline thin films by recording Rocking curves of the dominant Ta Bragg peaks \cite{blanton_use_1991}, see \autoref{fig:fig2}e. To this end, the detector is fixed, while the incident angle $\omega$ is varied with respect to the sample, revealing detrimental effects such as variations in the crystal orientation or film strain as broadening of the Rocking curve. Normalized data are shown in \autoref{fig:fig2}e, including fits to Pseudo-Voigt profiles from which fitted peak widths can be extracted as metric for film quality. While the $\beta$-Ta film exhibits the widest Rocking curve, we find that the width of the TaN-Ta related Rocking curve exceeds that of the Nb-Ta film by more than 3-fold. This indication of inferior crystal quality of the bare- and TaN-Ta films is in agreement with the superior microwave performance of the Nb-Ta resonators.

\begin{figure*}[th!]
\centering
    \includegraphics[scale=1]{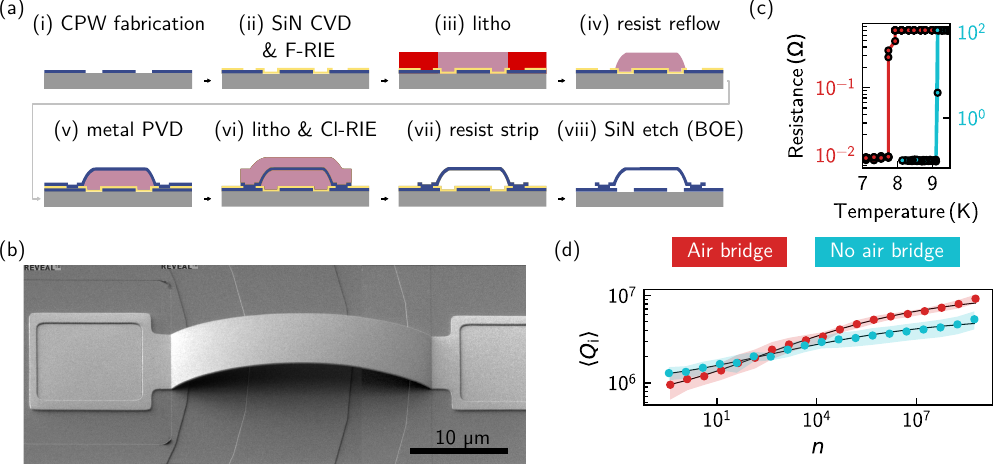}
    \caption{\fontsize{10pt}{12pt}\selectfont\textbf{Nb resonators with air bridges} (a) Schematic of air bridge fabrication process. Steps include dielectric encapsulation of pre-patterned planar structure and opening of air bridge contact area (i)-(ii), resist reflow process (iii)-(iv), Nb deposition and structuring of air bridges (v)-(vi) and removal of resist and dielectric layers (vii)-(viii). (b) SEM image of Nb air bridge crossing center conductor of CPW resonator. (c) Resistance of planar, patterned Nb film (cyan) and air bridge test structure (red) in dependence of temperature. Discontinuity below \qty{7.7}{K} indicates air bridge transition to superconducting state. (d) Cryogenic performance of resonators bridged by up to \num{35} Nb air bridges. Air bridge resonator data comprises measurements of 12 individual resonator structures, reference data is identical to cyan trace from \autoref{fig:fig1}a. Q-factor power-dependence of resonators including air bridges is similar to the reference group. Especially for higher photon numbers, $Q_\mathrm{i}$ appears to be improved on air bridged resonator samples. In the low photon number regime, the reference sample is slightly superior, probably due to dielectric loss caused by fabrication-related residues in the air bridge sample.}
    \label{fig:air bridge}
\end{figure*}
\section{Nb air bridges.}
In addition, we have developed an industry-compatible fabrication process to form Nb air bridges. Such elements are a widely adopted in superconducting quantum circuits to achieve interconnections between disjoint chip regions \cite{krinner_realizing_2022,valles-sanclemente_post-fabrication_2023}, but have been rarely implemented on a Nb material platform \cite{schicke_integrated_2003,wan_fabrication_2024,bruckmoser_niobium_2025}.

\paragraph{Fabrication.}
Wafers patterned according to the process illustrated in \autoref{fig:fig1}a are encapsulated with a \qty{80}{\nm} SiN passivation layer, deposited via plasma-enhanced CVD (see \autoref{fig:air bridge}a.) 
Using i-Line lithography and fluor-based RIE, we define and etch contact areas to connect air bridges to the metalization layer underneath. Subsequently, the wafer is coated with a layer of photoresist, which is subjected to a reflow process after exposure and development. This step creates the arc profile and serves to enhance the mechanical stability of the air bridges. Afterwards, another \qty{150}{\nm} Nb film is sputtered on the wafer and subsequently patterned into air bridges. After wet chemical removal of the photoresist layers, both above and underneath the bridges, the dielectric is finally removed via immersion in a BOE soluton at \qty{40}{\degreeCelsius}. A SEM image of a finalized air bridge is shown in \autoref{fig:air bridge}b.

Before electrical characterization, we optically inspect fully processed wafers for damaged or removed air bridges. To this end, we analyze six CPW resonator samples picked at random from wafer center, half-radius and edge across two wafers, containing \num{368} air bridges each. We observe no indication of missing air bridges and find all air bridges intact.

\paragraph{Electrical characterization of air bridges.}
Testing the superconducting properties of the air bridges, we perform transport measurements between Nb pads connected via \num{20} air bridges each. Sweeping the temperature at the mixing chamber stage of our cryostat, we screen the resistance across the sample. We observe a decrease of resistance over two orders of magnitude at a temperature slightly below \qty{8}{K}, indicating a transition to the superconducting state (see \autoref{fig:air bridge}c). The transition consists of two discontinuities (\autoref{fig:air bridge}c, red data), including a jump to an intermediate resistance level below \qty{7.9}{K} and a second decrease to the noise floor at \qty{7.7}{K}. We associate these with the film and the air bridges turning superconducting, respectively. 
For comparison, we measure $T_\mathrm{c}$ of a bare Nb thin film (\autoref{fig:air bridge}c, cyan data), consisting only of an unpatterned metalization layer on Si, which we find to be in good agreement with the bulk $T_\mathrm{c}=\qty{9.2}{\Kelvin}$ \cite{peiniger_superconducting_1985}. Therefore, we attribute the here observed lower $T_\mathrm{c}$ of the air bridge sample to the processing overhead of forming the air bridges rather than to its thin film nature. 
\\
To assess the impact of the air bridge process on losses in the microwave regime, we compare the quality factors of CPW resonators with the center conductor crossed by air bridges (see \autoref{fig:air bridge}b) to Nb resonators fabricated according to the process described in \autoref{fig:fig1}a, after an additional post-processing BOE dip. We observe an overall improvement of $\langle Q_\mathrm{i}\rangle$ across large photon populations $\langle n\rangle>100$ on the sample including air bridges, as depicted in \autoref{fig:air bridge}d. In contrast, in the small $\langle n\rangle$-regime, the reference sample performs slightly better. Accordingly, fit results (compare \autoref{t1}) indicate an increased TLS loss contribution $F\delta_\mathrm{TLS}^0$ and a smaller power-independent loss $\delta_0$. We hypothesize that the improved high power behavior may be accounted for by the air bridges breaking otherwise present, parasitic chip or slotline modes, whereas we attribute the reduced performance in the low power regime to the additional dielectric loss due to resist or SiN residues from the intermediate encapsulation.

\section{Conclusion and outlook}
We have demonstrated the fabrication of high-quality superconducting quantum circuits, using an industrial \qty{200}{\mm} semiconductor production line. We achieved high internal quality factors, on average exceeding $10^6$ in the single-photon regime, on both Nb- and Ta-based resonators, and have reported on the sucessful fabrication of superconducting Nb air bridges. Having verified the basic suitability of industrial semiconductor tools for the fabrication of superconducting quantum circuits paves the way towards leveraging decades of expertise in process control and stability acquired by semiconductor industry. Future work will focus on integrating Josephson junctions to build larger, more complex circuits in a reproducible and scalable manner, putting our capabilities in terms of reproducibility and parameter spread to the test. In order to make circuits truly scalable, we are additionally exploring ways to build a compact, multi-layer superconducting circuit technology that avoids the typical increase in dielectric loss when going to more compact circuit designs.\\

\begin{acknowledgments}
We thank S. Boguth for supporting during mask data production. We thank K. Aufinger from Infineon's technical development team. We thank B. Endres, S. Lipps and S. Henneck from Infineon's UPD department. We thank F. Haslbeck for support with the $T_\mathrm{c}$ measurements. We further thank M. Müller and S. Geprägs for support with the XRD measurements. We thank Infineon's failure analysis department for the support with post-processing BOE treatments of resonator samples, as well as N. Bruckmoser, who additionally supported with the development measurement scripts. Finally, we thank L. Koch for fruitful discussions about fabrication techniques for superconducting resonators. This research was partly supported by the German Federal Ministry of Research, Technology and Space (BMFTR) via the MUNIQC-SC (13N16183) and GeQCoS (13N15682) funding projects.
\end{acknowledgments}

\appendix

\bibliography{bibliography}

\begin{thebibliography}{30}%
\makeatletter
\providecommand \@ifxundefined [1]{%
 \@ifx{#1\undefined}
}%
\providecommand \@ifnum [1]{%
 \ifnum #1\expandafter \@firstoftwo
 \else \expandafter \@secondoftwo
 \fi
}%
\providecommand \@ifx [1]{%
 \ifx #1\expandafter \@firstoftwo
 \else \expandafter \@secondoftwo
 \fi
}%
\providecommand \natexlab [1]{#1}%
\providecommand \enquote  [1]{``#1''}%
\providecommand \bibnamefont  [1]{#1}%
\providecommand \bibfnamefont [1]{#1}%
\providecommand \citenamefont [1]{#1}%
\providecommand \href@noop [0]{\@secondoftwo}%
\providecommand \href [0]{\begingroup \@sanitize@url \@href}%
\providecommand \@href[1]{\@@startlink{#1}\@@href}%
\providecommand \@@href[1]{\endgroup#1\@@endlink}%
\providecommand \@sanitize@url [0]{\catcode `\\12\catcode `\$12\catcode
  `\&12\catcode `\#12\catcode `\^12\catcode `\_12\catcode `\%12\relax}%
\providecommand \@@startlink[1]{}%
\providecommand \@@endlink[0]{}%
\providecommand \url  [0]{\begingroup\@sanitize@url \@url }%
\providecommand \@url [1]{\endgroup\@href {#1}{\urlprefix }}%
\providecommand \urlprefix  [0]{URL }%
\providecommand \Eprint [0]{\href }%
\providecommand \doibase [0]{https://doi.org/}%
\providecommand \selectlanguage [0]{\@gobble}%
\providecommand \bibinfo  [0]{\@secondoftwo}%
\providecommand \bibfield  [0]{\@secondoftwo}%
\providecommand \translation [1]{[#1]}%
\providecommand \BibitemOpen [0]{}%
\providecommand \bibitemStop [0]{}%
\providecommand \bibitemNoStop [0]{.\EOS\space}%
\providecommand \EOS [0]{\spacefactor3000\relax}%
\providecommand \BibitemShut  [1]{\csname bibitem#1\endcsname}%
\let\auto@bib@innerbib\@empty
\bibitem [{\citenamefont {Kjaergaard}\ \emph {et~al.}(2020)\citenamefont
  {Kjaergaard}, \citenamefont {Schwartz}, \citenamefont {Braum\"{u}ller},
  \citenamefont {Krantz}, \citenamefont {Wang}, \citenamefont {Gustavsson},\
  and\ \citenamefont {Oliver}}]{kjaergaard_superconducting_2020}%
  \BibitemOpen
  \bibfield  {author} {\bibinfo {author} {\bibfnamefont {M.}~\bibnamefont
  {Kjaergaard}}, \bibinfo {author} {\bibfnamefont {M.~E.}\ \bibnamefont
  {Schwartz}}, \bibinfo {author} {\bibfnamefont {J.}~\bibnamefont
  {Braum\"{u}ller}}, \bibinfo {author} {\bibfnamefont {P.}~\bibnamefont
  {Krantz}}, \bibinfo {author} {\bibfnamefont {J.~I.-J.}\ \bibnamefont {Wang}},
  \bibinfo {author} {\bibfnamefont {S.}~\bibnamefont {Gustavsson}},\ and\
  \bibinfo {author} {\bibfnamefont {W.~D.}\ \bibnamefont {Oliver}},\ }\bibfield
   {title} {\bibinfo {title} {Superconducting {Qubits}: {Current} {State} of
  {Play}},\ }\href
  {https://www.annualreviews.org/doi/10.1146/annurev-conmatphys-031119-050605}
  {\bibfield  {journal} {\bibinfo  {journal} {Annual Review of Condensed Matter
  Physics}\ }\textbf {\bibinfo {volume} {11}},\ \bibinfo {pages} {369}
  (\bibinfo {year} {2020})}\BibitemShut {NoStop}%
\bibitem [{\citenamefont {Mohseni}\ \emph {et~al.}(2024)\citenamefont
  {Mohseni}, \citenamefont {Scherer}, \citenamefont {Johnson}, \citenamefont
  {Wertheim}, \citenamefont {Otten}, \citenamefont {Aadit}, \citenamefont
  {Bresniker}, \citenamefont {Camsari}, \citenamefont {Chapman}, \citenamefont
  {Chatterjee}, \citenamefont {Dagnew}, \citenamefont {Esposito}, \citenamefont
  {Fahim}, \citenamefont {Fiorentino}, \citenamefont {Khalid}, \citenamefont
  {Kong}, \citenamefont {Kulchytskyy}, \citenamefont {Li}, \citenamefont
  {Lott}, \citenamefont {Markov}, \citenamefont {McDermott}, \citenamefont
  {Pedretti}, \citenamefont {Gajjar}, \citenamefont {Silva}, \citenamefont
  {Sorebo}, \citenamefont {Spentzouris}, \citenamefont {Steiner}, \citenamefont
  {Torosov}, \citenamefont {Venturelli}, \citenamefont {Visser}, \citenamefont
  {Webb}, \citenamefont {Zhan}, \citenamefont {Cohen}, \citenamefont {Ronagh},
  \citenamefont {Ho}, \citenamefont {Beausoleil},\ and\ \citenamefont
  {Martinis}}]{mohseni_how_2024}%
  \BibitemOpen
  \bibfield  {author} {\bibinfo {author} {\bibfnamefont {M.}~\bibnamefont
  {Mohseni}}, \bibinfo {author} {\bibfnamefont {A.}~\bibnamefont {Scherer}},
  \bibinfo {author} {\bibfnamefont {K.~G.}\ \bibnamefont {Johnson}}, \bibinfo
  {author} {\bibfnamefont {O.}~\bibnamefont {Wertheim}}, \bibinfo {author}
  {\bibfnamefont {M.}~\bibnamefont {Otten}}, \bibinfo {author} {\bibfnamefont
  {N.~A.}\ \bibnamefont {Aadit}}, \bibinfo {author} {\bibfnamefont {K.~M.}\
  \bibnamefont {Bresniker}}, \bibinfo {author} {\bibfnamefont {K.~Y.}\
  \bibnamefont {Camsari}}, \bibinfo {author} {\bibfnamefont {B.}~\bibnamefont
  {Chapman}}, \bibinfo {author} {\bibfnamefont {S.}~\bibnamefont {Chatterjee}},
  \bibinfo {author} {\bibfnamefont {G.~A.}\ \bibnamefont {Dagnew}}, \bibinfo
  {author} {\bibfnamefont {A.}~\bibnamefont {Esposito}}, \bibinfo {author}
  {\bibfnamefont {F.}~\bibnamefont {Fahim}}, \bibinfo {author} {\bibfnamefont
  {M.}~\bibnamefont {Fiorentino}}, \bibinfo {author} {\bibfnamefont
  {A.}~\bibnamefont {Khalid}}, \bibinfo {author} {\bibfnamefont
  {X.}~\bibnamefont {Kong}}, \bibinfo {author} {\bibfnamefont {B.}~\bibnamefont
  {Kulchytskyy}}, \bibinfo {author} {\bibfnamefont {R.}~\bibnamefont {Li}},
  \bibinfo {author} {\bibfnamefont {P.~A.}\ \bibnamefont {Lott}}, \bibinfo
  {author} {\bibfnamefont {I.~L.}\ \bibnamefont {Markov}}, \bibinfo {author}
  {\bibfnamefont {R.~F.}\ \bibnamefont {McDermott}}, \bibinfo {author}
  {\bibfnamefont {G.}~\bibnamefont {Pedretti}}, \bibinfo {author}
  {\bibfnamefont {A.}~\bibnamefont {Gajjar}}, \bibinfo {author} {\bibfnamefont
  {A.}~\bibnamefont {Silva}}, \bibinfo {author} {\bibfnamefont
  {J.}~\bibnamefont {Sorebo}}, \bibinfo {author} {\bibfnamefont
  {P.}~\bibnamefont {Spentzouris}}, \bibinfo {author} {\bibfnamefont
  {Z.}~\bibnamefont {Steiner}}, \bibinfo {author} {\bibfnamefont
  {B.}~\bibnamefont {Torosov}}, \bibinfo {author} {\bibfnamefont
  {D.}~\bibnamefont {Venturelli}}, \bibinfo {author} {\bibfnamefont {R.~J.}\
  \bibnamefont {Visser}}, \bibinfo {author} {\bibfnamefont {Z.}~\bibnamefont
  {Webb}}, \bibinfo {author} {\bibfnamefont {X.}~\bibnamefont {Zhan}}, \bibinfo
  {author} {\bibfnamefont {Y.}~\bibnamefont {Cohen}}, \bibinfo {author}
  {\bibfnamefont {P.}~\bibnamefont {Ronagh}}, \bibinfo {author} {\bibfnamefont
  {A.}~\bibnamefont {Ho}}, \bibinfo {author} {\bibfnamefont {R.~G.}\
  \bibnamefont {Beausoleil}},\ and\ \bibinfo {author} {\bibfnamefont {J.~M.}\
  \bibnamefont {Martinis}},\ }\href {http://arxiv.org/abs/2411.10406} {\bibinfo
  {title} {How to {Build} a {Quantum} {Supercomputer}: {Scaling} {Challenges}
  and {Opportunities}}} (\bibinfo {year} {2024}),\ \bibinfo {note}
  {arXiv:2411.10406 [quant-ph]}\BibitemShut {NoStop}%
\bibitem [{\citenamefont {Van~Damme}\ \emph {et~al.}(2024)\citenamefont
  {Van~Damme}, \citenamefont {Massar}, \citenamefont {Acharya}, \citenamefont
  {Ivanov}, \citenamefont {Perez~Lozano}, \citenamefont {Canvel}, \citenamefont
  {Demarets}, \citenamefont {Vangoidsenhoven}, \citenamefont {Hermans},
  \citenamefont {Lai}, \citenamefont {Vadiraj}, \citenamefont {Mongillo},
  \citenamefont {Wan}, \citenamefont {De~Boeck}, \citenamefont {Potočnik},\
  and\ \citenamefont {De~Greve}}]{van_damme_advanced_2024}%
  \BibitemOpen
  \bibfield  {author} {\bibinfo {author} {\bibfnamefont {J.}~\bibnamefont
  {Van~Damme}}, \bibinfo {author} {\bibfnamefont {S.}~\bibnamefont {Massar}},
  \bibinfo {author} {\bibfnamefont {R.}~\bibnamefont {Acharya}}, \bibinfo
  {author} {\bibfnamefont {T.}~\bibnamefont {Ivanov}}, \bibinfo {author}
  {\bibfnamefont {D.}~\bibnamefont {Perez~Lozano}}, \bibinfo {author}
  {\bibfnamefont {Y.}~\bibnamefont {Canvel}}, \bibinfo {author} {\bibfnamefont
  {M.}~\bibnamefont {Demarets}}, \bibinfo {author} {\bibfnamefont
  {D.}~\bibnamefont {Vangoidsenhoven}}, \bibinfo {author} {\bibfnamefont
  {Y.}~\bibnamefont {Hermans}}, \bibinfo {author} {\bibfnamefont {J.~G.}\
  \bibnamefont {Lai}}, \bibinfo {author} {\bibfnamefont {A.~M.}\ \bibnamefont
  {Vadiraj}}, \bibinfo {author} {\bibfnamefont {M.}~\bibnamefont {Mongillo}},
  \bibinfo {author} {\bibfnamefont {D.}~\bibnamefont {Wan}}, \bibinfo {author}
  {\bibfnamefont {J.}~\bibnamefont {De~Boeck}}, \bibinfo {author}
  {\bibfnamefont {A.}~\bibnamefont {Potočnik}},\ and\ \bibinfo {author}
  {\bibfnamefont {K.}~\bibnamefont {De~Greve}},\ }\bibfield  {title} {\bibinfo
  {title} {Advanced {CMOS} manufacturing of superconducting qubits on 300 mm
  wafers},\ }\href {https://www.nature.com/articles/s41586-024-07941-9}
  {\bibfield  {journal} {\bibinfo  {journal} {Nature}\ }\textbf {\bibinfo
  {volume} {634}},\ \bibinfo {pages} {74} (\bibinfo {year} {2024})}\BibitemShut
  {NoStop}%
\bibitem [{\citenamefont {Blais}\ \emph {et~al.}(2004)\citenamefont {Blais},
  \citenamefont {Huang}, \citenamefont {Wallraff}, \citenamefont {Girvin},\
  and\ \citenamefont {Schoelkopf}}]{blais_cavity_2004}%
  \BibitemOpen
  \bibfield  {author} {\bibinfo {author} {\bibfnamefont {A.}~\bibnamefont
  {Blais}}, \bibinfo {author} {\bibfnamefont {R.-S.}\ \bibnamefont {Huang}},
  \bibinfo {author} {\bibfnamefont {A.}~\bibnamefont {Wallraff}}, \bibinfo
  {author} {\bibfnamefont {S.~M.}\ \bibnamefont {Girvin}},\ and\ \bibinfo
  {author} {\bibfnamefont {R.~J.}\ \bibnamefont {Schoelkopf}},\ }\bibfield
  {title} {\bibinfo {title} {Cavity quantum electrodynamics for superconducting
  electrical circuits: {An} architecture for quantum computation},\ }\href
  {https://link.aps.org/doi/10.1103/PhysRevA.69.062320} {\bibfield  {journal}
  {\bibinfo  {journal} {Physical Review A}\ }\textbf {\bibinfo {volume} {69}},\
  \bibinfo {pages} {062320} (\bibinfo {year} {2004})}\BibitemShut {NoStop}%
\bibitem [{\citenamefont {Wallraff}\ \emph {et~al.}(2004)\citenamefont
  {Wallraff}, \citenamefont {Schuster}, \citenamefont {Blais}, \citenamefont
  {Frunzio}, \citenamefont {Huang}, \citenamefont {Majer}, \citenamefont
  {Kumar}, \citenamefont {Girvin},\ and\ \citenamefont
  {Schoelkopf}}]{wallraff_strong_2004}%
  \BibitemOpen
  \bibfield  {author} {\bibinfo {author} {\bibfnamefont {A.}~\bibnamefont
  {Wallraff}}, \bibinfo {author} {\bibfnamefont {D.~I.}\ \bibnamefont
  {Schuster}}, \bibinfo {author} {\bibfnamefont {A.}~\bibnamefont {Blais}},
  \bibinfo {author} {\bibfnamefont {L.}~\bibnamefont {Frunzio}}, \bibinfo
  {author} {\bibfnamefont {R.-S.}\ \bibnamefont {Huang}}, \bibinfo {author}
  {\bibfnamefont {J.}~\bibnamefont {Majer}}, \bibinfo {author} {\bibfnamefont
  {S.}~\bibnamefont {Kumar}}, \bibinfo {author} {\bibfnamefont {S.~M.}\
  \bibnamefont {Girvin}},\ and\ \bibinfo {author} {\bibfnamefont {R.~J.}\
  \bibnamefont {Schoelkopf}},\ }\bibfield  {title} {\bibinfo {title} {Strong
  coupling of a single photon to a superconducting qubit using circuit quantum
  electrodynamics},\ }\href {https://www.nature.com/articles/nature02851}
  {\bibfield  {journal} {\bibinfo  {journal} {Nature}\ }\textbf {\bibinfo
  {volume} {431}},\ \bibinfo {pages} {162} (\bibinfo {year}
  {2004})}\BibitemShut {NoStop}%
\bibitem [{\citenamefont {Martinis}\ \emph {et~al.}(2005)\citenamefont
  {Martinis}, \citenamefont {Cooper}, \citenamefont {McDermott}, \citenamefont
  {Steffen}, \citenamefont {Ansmann}, \citenamefont {Osborn}, \citenamefont
  {Cicak}, \citenamefont {Oh}, \citenamefont {Pappas}, \citenamefont
  {Simmonds},\ and\ \citenamefont {Yu}}]{martinis_decoherence_2005}%
  \BibitemOpen
  \bibfield  {author} {\bibinfo {author} {\bibfnamefont {J.~M.}\ \bibnamefont
  {Martinis}}, \bibinfo {author} {\bibfnamefont {K.~B.}\ \bibnamefont
  {Cooper}}, \bibinfo {author} {\bibfnamefont {R.}~\bibnamefont {McDermott}},
  \bibinfo {author} {\bibfnamefont {M.}~\bibnamefont {Steffen}}, \bibinfo
  {author} {\bibfnamefont {M.}~\bibnamefont {Ansmann}}, \bibinfo {author}
  {\bibfnamefont {K.~D.}\ \bibnamefont {Osborn}}, \bibinfo {author}
  {\bibfnamefont {K.}~\bibnamefont {Cicak}}, \bibinfo {author} {\bibfnamefont
  {S.}~\bibnamefont {Oh}}, \bibinfo {author} {\bibfnamefont {D.~P.}\
  \bibnamefont {Pappas}}, \bibinfo {author} {\bibfnamefont {R.~W.}\
  \bibnamefont {Simmonds}},\ and\ \bibinfo {author} {\bibfnamefont {C.~C.}\
  \bibnamefont {Yu}},\ }\bibfield  {title} {\bibinfo {title} {Decoherence in
  {Josephson} {Qubits} from {Dielectric} {Loss}},\ }\href
  {https://link.aps.org/doi/10.1103/PhysRevLett.95.210503} {\bibfield
  {journal} {\bibinfo  {journal} {Physical Review Letters}\ }\textbf {\bibinfo
  {volume} {95}},\ \bibinfo {pages} {210503} (\bibinfo {year}
  {2005})}\BibitemShut {NoStop}%
\bibitem [{\citenamefont {McRae}\ \emph {et~al.}(2020)\citenamefont {McRae},
  \citenamefont {Wang}, \citenamefont {Gao}, \citenamefont {Vissers},
  \citenamefont {Brecht}, \citenamefont {Dunsworth}, \citenamefont {Pappas},\
  and\ \citenamefont {Mutus}}]{mcrae_materials_2020}%
  \BibitemOpen
  \bibfield  {author} {\bibinfo {author} {\bibfnamefont {C.~R.~H.}\
  \bibnamefont {McRae}}, \bibinfo {author} {\bibfnamefont {H.}~\bibnamefont
  {Wang}}, \bibinfo {author} {\bibfnamefont {J.}~\bibnamefont {Gao}}, \bibinfo
  {author} {\bibfnamefont {M.~R.}\ \bibnamefont {Vissers}}, \bibinfo {author}
  {\bibfnamefont {T.}~\bibnamefont {Brecht}}, \bibinfo {author} {\bibfnamefont
  {A.}~\bibnamefont {Dunsworth}}, \bibinfo {author} {\bibfnamefont {D.~P.}\
  \bibnamefont {Pappas}},\ and\ \bibinfo {author} {\bibfnamefont
  {J.}~\bibnamefont {Mutus}},\ }\bibfield  {title} {\bibinfo {title} {Materials
  loss measurements using superconducting microwave resonators},\ }\href
  {https://pubs.aip.org/rsi/article/91/9/091101/906092/Materials-loss-measurements-using-superconducting}
  {\bibfield  {journal} {\bibinfo  {journal} {Review of Scientific
  Instruments}\ }\textbf {\bibinfo {volume} {91}},\ \bibinfo {pages} {091101}
  (\bibinfo {year} {2020})}\BibitemShut {NoStop}%
\bibitem [{\citenamefont {Schicke}\ and\ \citenamefont
  {Schuster}(2003)}]{schicke_integrated_2003}%
  \BibitemOpen
  \bibfield  {author} {\bibinfo {author} {\bibfnamefont {M.}~\bibnamefont
  {Schicke}}\ and\ \bibinfo {author} {\bibfnamefont {K.}~\bibnamefont
  {Schuster}},\ }\bibfield  {title} {\bibinfo {title} {Integrated niobium thin
  film air bridges as variable capacitors for superconducting {GHz} electronic
  circuits},\ }\href {https://ieeexplore.ieee.org/document/1211560/} {\bibfield
   {journal} {\bibinfo  {journal} {IEEE Transactions on Applied
  Superconductivity}\ }\textbf {\bibinfo {volume} {13}},\ \bibinfo {pages}
  {135} (\bibinfo {year} {2003})}\BibitemShut {NoStop}%
\bibitem [{\citenamefont {Wan}\ \emph {et~al.}(2024)\citenamefont {Wan},
  \citenamefont {Mongillo}, \citenamefont {Canvel}, \citenamefont {Lozano},
  \citenamefont {Tobback}, \citenamefont {Ivanov}, \citenamefont {Pacco},
  \citenamefont {Piao}, \citenamefont {Massar}, \citenamefont {Potočnik},\
  and\ \citenamefont {De~Greve}}]{wan_fabrication_2024}%
  \BibitemOpen
  \bibfield  {author} {\bibinfo {author} {\bibfnamefont {D.}~\bibnamefont
  {Wan}}, \bibinfo {author} {\bibfnamefont {M.}~\bibnamefont {Mongillo}},
  \bibinfo {author} {\bibfnamefont {Y.}~\bibnamefont {Canvel}}, \bibinfo
  {author} {\bibfnamefont {D.~P.}\ \bibnamefont {Lozano}}, \bibinfo {author}
  {\bibfnamefont {B.}~\bibnamefont {Tobback}}, \bibinfo {author} {\bibfnamefont
  {T.}~\bibnamefont {Ivanov}}, \bibinfo {author} {\bibfnamefont
  {A.}~\bibnamefont {Pacco}}, \bibinfo {author} {\bibfnamefont
  {X.}~\bibnamefont {Piao}}, \bibinfo {author} {\bibfnamefont {S.}~\bibnamefont
  {Massar}}, \bibinfo {author} {\bibfnamefont {A.}~\bibnamefont {Potočnik}},\
  and\ \bibinfo {author} {\bibfnamefont {K.}~\bibnamefont {De~Greve}},\
  }\bibfield  {title} {\bibinfo {title} {Fabrication of {Superconducting} {Nb}
  {Airbridges} in a 300 mm {Pilot} {Line} for {Quantum} {Technologies}},\ }in\
  \href {https://ieeexplore.ieee.org/document/10732203/} {\emph {\bibinfo
  {booktitle} {2024 {IEEE} {International} {Interconnect} {Technology}
  {Conference} ({IITC})}}}\ (\bibinfo {year} {2024})\ pp.\ \bibinfo {pages}
  {1--3},\ \bibinfo {note} {iSSN: 2380-6338}\BibitemShut {NoStop}%
\bibitem [{\citenamefont {Bruckmoser}\ \emph {et~al.}(2025)\citenamefont
  {Bruckmoser}, \citenamefont {Koch}, \citenamefont {Tsitsilin}, \citenamefont
  {Grammer}, \citenamefont {Bunch}, \citenamefont {Richard}, \citenamefont
  {Schirk}, \citenamefont {Wallner}, \citenamefont {Feigl}, \citenamefont
  {Schneider}, \citenamefont {Geprägs}, \citenamefont {Bader}, \citenamefont
  {Althammer}, \citenamefont {Södergren},\ and\ \citenamefont
  {Filipp}}]{bruckmoser_niobium_2025}%
  \BibitemOpen
  \bibfield  {author} {\bibinfo {author} {\bibfnamefont {N.}~\bibnamefont
  {Bruckmoser}}, \bibinfo {author} {\bibfnamefont {L.}~\bibnamefont {Koch}},
  \bibinfo {author} {\bibfnamefont {I.}~\bibnamefont {Tsitsilin}}, \bibinfo
  {author} {\bibfnamefont {M.}~\bibnamefont {Grammer}}, \bibinfo {author}
  {\bibfnamefont {D.}~\bibnamefont {Bunch}}, \bibinfo {author} {\bibfnamefont
  {L.}~\bibnamefont {Richard}}, \bibinfo {author} {\bibfnamefont
  {J.}~\bibnamefont {Schirk}}, \bibinfo {author} {\bibfnamefont
  {F.}~\bibnamefont {Wallner}}, \bibinfo {author} {\bibfnamefont
  {J.}~\bibnamefont {Feigl}}, \bibinfo {author} {\bibfnamefont {C.~M.~F.}\
  \bibnamefont {Schneider}}, \bibinfo {author} {\bibfnamefont {S.}~\bibnamefont
  {Geprägs}}, \bibinfo {author} {\bibfnamefont {V.~P.}\ \bibnamefont {Bader}},
  \bibinfo {author} {\bibfnamefont {M.}~\bibnamefont {Althammer}}, \bibinfo
  {author} {\bibfnamefont {L.}~\bibnamefont {Södergren}},\ and\ \bibinfo
  {author} {\bibfnamefont {S.}~\bibnamefont {Filipp}},\ }\href
  {http://arxiv.org/abs/2503.12076} {\bibinfo {title} {Niobium {Air} {Bridges}
  as a {Low}-{Loss} {Component} for {Superconducting} {Quantum} {Hardware}}}
  (\bibinfo {year} {2025}),\ \bibinfo {note} {arXiv:2503.12076
  [quant-ph]}\BibitemShut {NoStop}%
\bibitem [{\citenamefont {Kern}(1990)}]{kern_evolution_1990}%
  \BibitemOpen
  \bibfield  {author} {\bibinfo {author} {\bibfnamefont {W.}~\bibnamefont
  {Kern}},\ }\bibfield  {title} {\bibinfo {title} {The {Evolution} of {Silicon}
  {Wafer} {Cleaning} {Technology}},\ }\href
  {https://iopscience.iop.org/article/10.1149/1.2086825/meta} {\bibfield
  {journal} {\bibinfo  {journal} {Journal of The Electrochemical Society}\
  }\textbf {\bibinfo {volume} {137}},\ \bibinfo {pages} {1887} (\bibinfo {year}
  {1990})},\ \bibinfo {note} {publisher: IOP Publishing}\BibitemShut {NoStop}%
\bibitem [{\citenamefont {Face}\ and\ \citenamefont
  {Prober}(1987)}]{face_nucleation_1987}%
  \BibitemOpen
  \bibfield  {author} {\bibinfo {author} {\bibfnamefont {D.~W.}\ \bibnamefont
  {Face}}\ and\ \bibinfo {author} {\bibfnamefont {D.~E.}\ \bibnamefont
  {Prober}},\ }\bibfield  {title} {\bibinfo {title} {Nucleation of
  body-centered-cubic tantalum films with a thin niobium underlayer},\ }\href
  {https://pubs.aip.org/jva/article/5/6/3408/246650/Nucleation-of-body-centered-cubic-tantalum-films}
  {\bibfield  {journal} {\bibinfo  {journal} {Journal of Vacuum Science \&
  Technology A: Vacuum, Surfaces, and Films}\ }\textbf {\bibinfo {volume}
  {5}},\ \bibinfo {pages} {3408} (\bibinfo {year} {1987})}\BibitemShut
  {NoStop}%
\bibitem [{\citenamefont {Gladczuk}\ \emph {et~al.}(2005)\citenamefont
  {Gladczuk}, \citenamefont {Patel}, \citenamefont {Demaree},\ and\
  \citenamefont {Sosnowski}}]{gladczuk_sputter_2005}%
  \BibitemOpen
  \bibfield  {author} {\bibinfo {author} {\bibfnamefont {L.}~\bibnamefont
  {Gladczuk}}, \bibinfo {author} {\bibfnamefont {A.}~\bibnamefont {Patel}},
  \bibinfo {author} {\bibfnamefont {J.~D.}\ \bibnamefont {Demaree}},\ and\
  \bibinfo {author} {\bibfnamefont {M.}~\bibnamefont {Sosnowski}},\ }\bibfield
  {title} {\bibinfo {title} {Sputter deposition of bcc tantalum films with
  {TaN} underlayers for protection of steel},\ }\href
  {https://linkinghub.elsevier.com/retrieve/pii/S0040609004014476} {\bibfield
  {journal} {\bibinfo  {journal} {Thin Solid Films}\ }\textbf {\bibinfo
  {volume} {476}},\ \bibinfo {pages} {295} (\bibinfo {year}
  {2005})}\BibitemShut {NoStop}%
\bibitem [{\citenamefont {Probst}\ \emph {et~al.}(2015)\citenamefont {Probst},
  \citenamefont {Song}, \citenamefont {Bushev}, \citenamefont {Ustinov},\ and\
  \citenamefont {Weides}}]{probst_efficient_2015}%
  \BibitemOpen
  \bibfield  {author} {\bibinfo {author} {\bibfnamefont {S.}~\bibnamefont
  {Probst}}, \bibinfo {author} {\bibfnamefont {F.~B.}\ \bibnamefont {Song}},
  \bibinfo {author} {\bibfnamefont {P.~A.}\ \bibnamefont {Bushev}}, \bibinfo
  {author} {\bibfnamefont {A.~V.}\ \bibnamefont {Ustinov}},\ and\ \bibinfo
  {author} {\bibfnamefont {M.}~\bibnamefont {Weides}},\ }\bibfield  {title}
  {\bibinfo {title} {Efficient and robust analysis of complex scattering data
  under noise in microwave resonators},\ }\href
  {https://pubs.aip.org/rsi/article/86/2/024706/360955/Efficient-and-robust-analysis-of-complex}
  {\bibfield  {journal} {\bibinfo  {journal} {Review of Scientific
  Instruments}\ }\textbf {\bibinfo {volume} {86}},\ \bibinfo {pages} {024706}
  (\bibinfo {year} {2015})}\BibitemShut {NoStop}%
\bibitem [{\citenamefont {Bruno}\ \emph {et~al.}(2015)\citenamefont {Bruno},
  \citenamefont {De~Lange}, \citenamefont {Asaad}, \citenamefont {Van
  Der~Enden}, \citenamefont {Langford},\ and\ \citenamefont
  {DiCarlo}}]{bruno_reducing_2015}%
  \BibitemOpen
  \bibfield  {author} {\bibinfo {author} {\bibfnamefont {A.}~\bibnamefont
  {Bruno}}, \bibinfo {author} {\bibfnamefont {G.}~\bibnamefont {De~Lange}},
  \bibinfo {author} {\bibfnamefont {S.}~\bibnamefont {Asaad}}, \bibinfo
  {author} {\bibfnamefont {K.~L.}\ \bibnamefont {Van Der~Enden}}, \bibinfo
  {author} {\bibfnamefont {N.~K.}\ \bibnamefont {Langford}},\ and\ \bibinfo
  {author} {\bibfnamefont {L.}~\bibnamefont {DiCarlo}},\ }\bibfield  {title}
  {\bibinfo {title} {Reducing intrinsic loss in superconducting resonators by
  surface treatment and deep etching of silicon substrates},\ }\href
  {https://pubs.aip.org/apl/article/106/18/182601/27784/Reducing-intrinsic-loss-in-superconducting}
  {\bibfield  {journal} {\bibinfo  {journal} {Applied Physics Letters}\
  }\textbf {\bibinfo {volume} {106}},\ \bibinfo {pages} {182601} (\bibinfo
  {year} {2015})}\BibitemShut {NoStop}%
\bibitem [{\citenamefont {Urade}\ \emph {et~al.}(2024)\citenamefont {Urade},
  \citenamefont {Yakushiji}, \citenamefont {Tsujimoto}, \citenamefont {Yamada},
  \citenamefont {Makise}, \citenamefont {Mizubayashi},\ and\ \citenamefont
  {Inomata}}]{urade_microwave_2024}%
  \BibitemOpen
  \bibfield  {author} {\bibinfo {author} {\bibfnamefont {Y.}~\bibnamefont
  {Urade}}, \bibinfo {author} {\bibfnamefont {K.}~\bibnamefont {Yakushiji}},
  \bibinfo {author} {\bibfnamefont {M.}~\bibnamefont {Tsujimoto}}, \bibinfo
  {author} {\bibfnamefont {T.}~\bibnamefont {Yamada}}, \bibinfo {author}
  {\bibfnamefont {K.}~\bibnamefont {Makise}}, \bibinfo {author} {\bibfnamefont
  {W.}~\bibnamefont {Mizubayashi}},\ and\ \bibinfo {author} {\bibfnamefont
  {K.}~\bibnamefont {Inomata}},\ }\bibfield  {title} {\bibinfo {title}
  {Microwave characterization of tantalum superconducting resonators on silicon
  substrate with niobium buffer layer},\ }\href
  {https://pubs.aip.org/apm/article/12/2/021132/3267941/Microwave-characterization-of-tantalum}
  {\bibfield  {journal} {\bibinfo  {journal} {APL Materials}\ }\textbf
  {\bibinfo {volume} {12}},\ \bibinfo {pages} {021132} (\bibinfo {year}
  {2024})}\BibitemShut {NoStop}%
\bibitem [{\citenamefont {Gao}(2008)}]{gao_physics_2008}%
  \BibitemOpen
  \bibfield  {author} {\bibinfo {author} {\bibfnamefont {J.}~\bibnamefont
  {Gao}},\ }\emph {\bibinfo {title} {The {Physics} of {Superconducting}
  {Microwave} {Resonators}}},\ \href@noop {} {Ph.D. thesis},\ \bibinfo
  {school} {California Institute of Technology}, \bibinfo {address} {Pasadena,
  CA} (\bibinfo {year} {2008})\BibitemShut {NoStop}%
\bibitem [{\citenamefont {Wang}\ \emph {et~al.}(2009)\citenamefont {Wang},
  \citenamefont {Hofheinz}, \citenamefont {Wenner}, \citenamefont {Ansmann},
  \citenamefont {Bialczak}, \citenamefont {Lenander}, \citenamefont {Lucero},
  \citenamefont {Neeley}, \citenamefont {O’Connell}, \citenamefont {Sank},
  \citenamefont {Weides}, \citenamefont {Cleland},\ and\ \citenamefont
  {Martinis}}]{wang_improving_2009}%
  \BibitemOpen
  \bibfield  {author} {\bibinfo {author} {\bibfnamefont {H.}~\bibnamefont
  {Wang}}, \bibinfo {author} {\bibfnamefont {M.}~\bibnamefont {Hofheinz}},
  \bibinfo {author} {\bibfnamefont {J.}~\bibnamefont {Wenner}}, \bibinfo
  {author} {\bibfnamefont {M.}~\bibnamefont {Ansmann}}, \bibinfo {author}
  {\bibfnamefont {R.~C.}\ \bibnamefont {Bialczak}}, \bibinfo {author}
  {\bibfnamefont {M.}~\bibnamefont {Lenander}}, \bibinfo {author}
  {\bibfnamefont {E.}~\bibnamefont {Lucero}}, \bibinfo {author} {\bibfnamefont
  {M.}~\bibnamefont {Neeley}}, \bibinfo {author} {\bibfnamefont {A.~D.}\
  \bibnamefont {O’Connell}}, \bibinfo {author} {\bibfnamefont
  {D.}~\bibnamefont {Sank}}, \bibinfo {author} {\bibfnamefont {M.}~\bibnamefont
  {Weides}}, \bibinfo {author} {\bibfnamefont {A.~N.}\ \bibnamefont
  {Cleland}},\ and\ \bibinfo {author} {\bibfnamefont {J.~M.}\ \bibnamefont
  {Martinis}},\ }\bibfield  {title} {\bibinfo {title} {Improving the coherence
  time of superconducting coplanar resonators},\ }\href
  {https://pubs.aip.org/apl/article/95/23/233508/120944/Improving-the-coherence-time-of-superconducting}
  {\bibfield  {journal} {\bibinfo  {journal} {Applied Physics Letters}\
  }\textbf {\bibinfo {volume} {95}},\ \bibinfo {pages} {233508} (\bibinfo
  {year} {2009})}\BibitemShut {NoStop}%
\bibitem [{\citenamefont {Crowley}\ \emph {et~al.}(2023)\citenamefont
  {Crowley}, \citenamefont {McLellan}, \citenamefont {Dutta}, \citenamefont
  {Shumiya}, \citenamefont {Place}, \citenamefont {Le}, \citenamefont {Gang},
  \citenamefont {Madhavan}, \citenamefont {Bland}, \citenamefont {Chang},
  \citenamefont {Khedkar}, \citenamefont {Feng}, \citenamefont {Umbarkar},
  \citenamefont {Gui}, \citenamefont {Rodgers}, \citenamefont {Jia},
  \citenamefont {Feldman}, \citenamefont {Lyon}, \citenamefont {Liu},
  \citenamefont {Cava}, \citenamefont {Houck},\ and\ \citenamefont
  {De~Leon}}]{crowley_disentangling_2023}%
  \BibitemOpen
  \bibfield  {author} {\bibinfo {author} {\bibfnamefont {K.~D.}\ \bibnamefont
  {Crowley}}, \bibinfo {author} {\bibfnamefont {R.~A.}\ \bibnamefont
  {McLellan}}, \bibinfo {author} {\bibfnamefont {A.}~\bibnamefont {Dutta}},
  \bibinfo {author} {\bibfnamefont {N.}~\bibnamefont {Shumiya}}, \bibinfo
  {author} {\bibfnamefont {A.~P.}\ \bibnamefont {Place}}, \bibinfo {author}
  {\bibfnamefont {X.~H.}\ \bibnamefont {Le}}, \bibinfo {author} {\bibfnamefont
  {Y.}~\bibnamefont {Gang}}, \bibinfo {author} {\bibfnamefont {T.}~\bibnamefont
  {Madhavan}}, \bibinfo {author} {\bibfnamefont {M.~P.}\ \bibnamefont {Bland}},
  \bibinfo {author} {\bibfnamefont {R.}~\bibnamefont {Chang}}, \bibinfo
  {author} {\bibfnamefont {N.}~\bibnamefont {Khedkar}}, \bibinfo {author}
  {\bibfnamefont {Y.~C.}\ \bibnamefont {Feng}}, \bibinfo {author}
  {\bibfnamefont {E.~A.}\ \bibnamefont {Umbarkar}}, \bibinfo {author}
  {\bibfnamefont {X.}~\bibnamefont {Gui}}, \bibinfo {author} {\bibfnamefont
  {L.~V.}\ \bibnamefont {Rodgers}}, \bibinfo {author} {\bibfnamefont
  {Y.}~\bibnamefont {Jia}}, \bibinfo {author} {\bibfnamefont {M.~M.}\
  \bibnamefont {Feldman}}, \bibinfo {author} {\bibfnamefont {S.~A.}\
  \bibnamefont {Lyon}}, \bibinfo {author} {\bibfnamefont {M.}~\bibnamefont
  {Liu}}, \bibinfo {author} {\bibfnamefont {R.~J.}\ \bibnamefont {Cava}},
  \bibinfo {author} {\bibfnamefont {A.~A.}\ \bibnamefont {Houck}},\ and\
  \bibinfo {author} {\bibfnamefont {N.~P.}\ \bibnamefont {De~Leon}},\
  }\bibfield  {title} {\bibinfo {title} {Disentangling {Losses} in {Tantalum}
  {Superconducting} {Circuits}},\ }\href
  {https://link.aps.org/doi/10.1103/PhysRevX.13.041005} {\bibfield  {journal}
  {\bibinfo  {journal} {Physical Review X}\ }\textbf {\bibinfo {volume} {13}},\
  \bibinfo {pages} {041005} (\bibinfo {year} {2023})}\BibitemShut {NoStop}%
\bibitem [{\citenamefont {Wisbey}\ \emph {et~al.}(2010)\citenamefont {Wisbey},
  \citenamefont {Gao}, \citenamefont {Vissers}, \citenamefont {da~Silva},
  \citenamefont {Kline}, \citenamefont {Vale},\ and\ \citenamefont
  {Pappas}}]{wisbey_effect_2010}%
  \BibitemOpen
  \bibfield  {author} {\bibinfo {author} {\bibfnamefont {D.~S.}\ \bibnamefont
  {Wisbey}}, \bibinfo {author} {\bibfnamefont {J.}~\bibnamefont {Gao}},
  \bibinfo {author} {\bibfnamefont {M.~R.}\ \bibnamefont {Vissers}}, \bibinfo
  {author} {\bibfnamefont {F.~C.~S.}\ \bibnamefont {da~Silva}}, \bibinfo
  {author} {\bibfnamefont {J.~S.}\ \bibnamefont {Kline}}, \bibinfo {author}
  {\bibfnamefont {L.}~\bibnamefont {Vale}},\ and\ \bibinfo {author}
  {\bibfnamefont {D.~P.}\ \bibnamefont {Pappas}},\ }\bibfield  {title}
  {\bibinfo {title} {Effect of metal/substrate interfaces on radio-frequency
  loss in superconducting coplanar waveguides},\ }\href
  {https://doi.org/10.1063/1.3499608} {\bibfield  {journal} {\bibinfo
  {journal} {Journal of Applied Physics}\ }\textbf {\bibinfo {volume} {108}},\
  \bibinfo {pages} {093918} (\bibinfo {year} {2010})}\BibitemShut {NoStop}%
\bibitem [{\citenamefont {Verjauw}\ \emph {et~al.}(2021)\citenamefont
  {Verjauw}, \citenamefont {Poto\v{c}nik}, \citenamefont {Mongillo},
  \citenamefont {Acharya}, \citenamefont {Mohiyaddin}, \citenamefont {Simion},
  \citenamefont {Pacco}, \citenamefont {Ivanov}, \citenamefont {Wan},
  \citenamefont {Vanleenhove}, \citenamefont {Souriau}, \citenamefont {Jussot},
  \citenamefont {Thiam}, \citenamefont {Swerts}, \citenamefont {Piao},
  \citenamefont {Couet}, \citenamefont {Heyns}, \citenamefont {Govoreanu},\
  and\ \citenamefont {Radu}}]{verjauw_investigation_2021}%
  \BibitemOpen
  \bibfield  {author} {\bibinfo {author} {\bibfnamefont {J.}~\bibnamefont
  {Verjauw}}, \bibinfo {author} {\bibfnamefont {A.}~\bibnamefont
  {Poto\v{c}nik}}, \bibinfo {author} {\bibfnamefont {M.}~\bibnamefont
  {Mongillo}}, \bibinfo {author} {\bibfnamefont {R.}~\bibnamefont {Acharya}},
  \bibinfo {author} {\bibfnamefont {F.}~\bibnamefont {Mohiyaddin}}, \bibinfo
  {author} {\bibfnamefont {G.}~\bibnamefont {Simion}}, \bibinfo {author}
  {\bibfnamefont {A.}~\bibnamefont {Pacco}}, \bibinfo {author} {\bibfnamefont
  {T.}~\bibnamefont {Ivanov}}, \bibinfo {author} {\bibfnamefont
  {D.}~\bibnamefont {Wan}}, \bibinfo {author} {\bibfnamefont {A.}~\bibnamefont
  {Vanleenhove}}, \bibinfo {author} {\bibfnamefont {L.}~\bibnamefont
  {Souriau}}, \bibinfo {author} {\bibfnamefont {J.}~\bibnamefont {Jussot}},
  \bibinfo {author} {\bibfnamefont {A.}~\bibnamefont {Thiam}}, \bibinfo
  {author} {\bibfnamefont {J.}~\bibnamefont {Swerts}}, \bibinfo {author}
  {\bibfnamefont {X.}~\bibnamefont {Piao}}, \bibinfo {author} {\bibfnamefont
  {S.}~\bibnamefont {Couet}}, \bibinfo {author} {\bibfnamefont
  {M.}~\bibnamefont {Heyns}}, \bibinfo {author} {\bibfnamefont
  {B.}~\bibnamefont {Govoreanu}},\ and\ \bibinfo {author} {\bibfnamefont
  {I.}~\bibnamefont {Radu}},\ }\bibfield  {title} {\bibinfo {title}
  {Investigation of {Microwave} {Loss} {Induced} by {Oxide} {Regrowth} in
  {High}- \textit{{Q}} {Niobium} {Resonators}},\ }\href
  {https://link.aps.org/doi/10.1103/PhysRevApplied.16.014018} {\bibfield
  {journal} {\bibinfo  {journal} {Physical Review Applied}\ }\textbf {\bibinfo
  {volume} {16}},\ \bibinfo {pages} {014018} (\bibinfo {year}
  {2021})}\BibitemShut {NoStop}%
\bibitem [{\citenamefont {Alto\'{e}}\ \emph {et~al.}(2022)\citenamefont
  {Alto\'{e}}, \citenamefont {Banerjee}, \citenamefont {Berk}, \citenamefont
  {Hajr}, \citenamefont {Schwartzberg}, \citenamefont {Song}, \citenamefont
  {Alghadeer}, \citenamefont {Aloni}, \citenamefont {Elowson}, \citenamefont
  {Kreikebaum}, \citenamefont {Wong}, \citenamefont {Griffin}, \citenamefont
  {Rao}, \citenamefont {Weber-Bargioni}, \citenamefont {Minor}, \citenamefont
  {Santiago}, \citenamefont {Cabrini}, \citenamefont {Siddiqi},\ and\
  \citenamefont {Ogletree}}]{altoe_localization_2022}%
  \BibitemOpen
  \bibfield  {author} {\bibinfo {author} {\bibfnamefont {M.~V.~P.}\
  \bibnamefont {Alto\'{e}}}, \bibinfo {author} {\bibfnamefont {A.}~\bibnamefont
  {Banerjee}}, \bibinfo {author} {\bibfnamefont {C.}~\bibnamefont {Berk}},
  \bibinfo {author} {\bibfnamefont {A.}~\bibnamefont {Hajr}}, \bibinfo {author}
  {\bibfnamefont {A.}~\bibnamefont {Schwartzberg}}, \bibinfo {author}
  {\bibfnamefont {C.}~\bibnamefont {Song}}, \bibinfo {author} {\bibfnamefont
  {M.}~\bibnamefont {Alghadeer}}, \bibinfo {author} {\bibfnamefont
  {S.}~\bibnamefont {Aloni}}, \bibinfo {author} {\bibfnamefont {M.~J.}\
  \bibnamefont {Elowson}}, \bibinfo {author} {\bibfnamefont {J.~M.}\
  \bibnamefont {Kreikebaum}}, \bibinfo {author} {\bibfnamefont {E.~K.}\
  \bibnamefont {Wong}}, \bibinfo {author} {\bibfnamefont {S.~M.}\ \bibnamefont
  {Griffin}}, \bibinfo {author} {\bibfnamefont {S.}~\bibnamefont {Rao}},
  \bibinfo {author} {\bibfnamefont {A.}~\bibnamefont {Weber-Bargioni}},
  \bibinfo {author} {\bibfnamefont {A.~M.}\ \bibnamefont {Minor}}, \bibinfo
  {author} {\bibfnamefont {D.~I.}\ \bibnamefont {Santiago}}, \bibinfo {author}
  {\bibfnamefont {S.}~\bibnamefont {Cabrini}}, \bibinfo {author} {\bibfnamefont
  {I.}~\bibnamefont {Siddiqi}},\ and\ \bibinfo {author} {\bibfnamefont {D.~F.}\
  \bibnamefont {Ogletree}},\ }\bibfield  {title} {\bibinfo {title}
  {Localization and {Mitigation} of {Loss} in {Niobium} {Superconducting}
  {Circuits}},\ }\href {https://link.aps.org/doi/10.1103/PRXQuantum.3.020312}
  {\bibfield  {journal} {\bibinfo  {journal} {PRX Quantum}\ }\textbf {\bibinfo
  {volume} {3}},\ \bibinfo {pages} {020312} (\bibinfo {year}
  {2022})}\BibitemShut {NoStop}%
\bibitem [{\citenamefont {M\"{u}ller}\ \emph {et~al.}(2019)\citenamefont
  {M\"{u}ller}, \citenamefont {Cole},\ and\ \citenamefont
  {Lisenfeld}}]{muller_towards_2019}%
  \BibitemOpen
  \bibfield  {author} {\bibinfo {author} {\bibfnamefont {C.}~\bibnamefont
  {M\"{u}ller}}, \bibinfo {author} {\bibfnamefont {J.~H.}\ \bibnamefont
  {Cole}},\ and\ \bibinfo {author} {\bibfnamefont {J.}~\bibnamefont
  {Lisenfeld}},\ }\bibfield  {title} {\bibinfo {title} {Towards understanding
  two-level-systems in amorphous solids: insights from quantum circuits},\
  }\href {https://iopscience.iop.org/article/10.1088/1361-6633/ab3a7e}
  {\bibfield  {journal} {\bibinfo  {journal} {Reports on Progress in Physics}\
  }\textbf {\bibinfo {volume} {82}},\ \bibinfo {pages} {124501} (\bibinfo
  {year} {2019})}\BibitemShut {NoStop}%
\bibitem [{\citenamefont {Anderson}\ \emph {et~al.}(1972)\citenamefont
  {Anderson}, \citenamefont {Halperin},\ and\ \citenamefont
  {Varma}}]{anderson_anomalous_1972}%
  \BibitemOpen
  \bibfield  {author} {\bibinfo {author} {\bibfnamefont {P.~w.}\ \bibnamefont
  {Anderson}}, \bibinfo {author} {\bibfnamefont {B.~I.}\ \bibnamefont
  {Halperin}},\ and\ \bibinfo {author} {\bibfnamefont {c.~M.}\ \bibnamefont
  {Varma}},\ }\bibfield  {title} {\bibinfo {title} {Anomalous low-temperature
  thermal properties of glasses and spin glasses},\ }\href
  {https://doi.org/10.1080/14786437208229210} {\bibfield  {journal} {\bibinfo
  {journal} {The Philosophical Magazine: A Journal of Theoretical Experimental
  and Applied Physics}\ }\textbf {\bibinfo {volume} {25}},\ \bibinfo {pages}
  {1} (\bibinfo {year} {1972})},\ \bibinfo {note} {publisher: Taylor \& Francis
  \_eprint: https://doi.org/10.1080/14786437208229210}\BibitemShut {NoStop}%
\bibitem [{\citenamefont {Phillips}(1987)}]{phillips_two-level_1987}%
  \BibitemOpen
  \bibfield  {author} {\bibinfo {author} {\bibfnamefont {W.~A.}\ \bibnamefont
  {Phillips}},\ }\bibfield  {title} {\bibinfo {title} {Two-level states in
  glasses},\ }\href {https://dx.doi.org/10.1088/0034-4885/50/12/003} {\bibfield
   {journal} {\bibinfo  {journal} {Reports on Progress in Physics}\ }\textbf
  {\bibinfo {volume} {50}},\ \bibinfo {pages} {1657} (\bibinfo {year}
  {1987})}\BibitemShut {NoStop}%
\bibitem [{\citenamefont {Dhundhwal}\ \emph {et~al.}(2025)\citenamefont
  {Dhundhwal}, \citenamefont {Duan}, \citenamefont {Brauch}, \citenamefont
  {Arabi}, \citenamefont {Fuchs}, \citenamefont {Haghighirad}, \citenamefont
  {Welle}, \citenamefont {Scharwaechter}, \citenamefont {Pal}, \citenamefont
  {Scheffler}, \citenamefont {Palomo}, \citenamefont {Leghtas}, \citenamefont
  {Murani}, \citenamefont {Hahn}, \citenamefont {Aghassi-Hagmann},
  \citenamefont {K\"{u}bel}, \citenamefont {Wulfhekel}, \citenamefont {Pop},\
  and\ \citenamefont {Reisinger}}]{dhundhwal_high_2025}%
  \BibitemOpen
  \bibfield  {author} {\bibinfo {author} {\bibfnamefont {R.}~\bibnamefont
  {Dhundhwal}}, \bibinfo {author} {\bibfnamefont {H.}~\bibnamefont {Duan}},
  \bibinfo {author} {\bibfnamefont {L.}~\bibnamefont {Brauch}}, \bibinfo
  {author} {\bibfnamefont {S.}~\bibnamefont {Arabi}}, \bibinfo {author}
  {\bibfnamefont {D.}~\bibnamefont {Fuchs}}, \bibinfo {author} {\bibfnamefont
  {A.-A.}\ \bibnamefont {Haghighirad}}, \bibinfo {author} {\bibfnamefont
  {A.}~\bibnamefont {Welle}}, \bibinfo {author} {\bibfnamefont
  {F.}~\bibnamefont {Scharwaechter}}, \bibinfo {author} {\bibfnamefont
  {S.}~\bibnamefont {Pal}}, \bibinfo {author} {\bibfnamefont {M.}~\bibnamefont
  {Scheffler}}, \bibinfo {author} {\bibfnamefont {J.}~\bibnamefont {Palomo}},
  \bibinfo {author} {\bibfnamefont {Z.}~\bibnamefont {Leghtas}}, \bibinfo
  {author} {\bibfnamefont {A.}~\bibnamefont {Murani}}, \bibinfo {author}
  {\bibfnamefont {H.}~\bibnamefont {Hahn}}, \bibinfo {author} {\bibfnamefont
  {J.}~\bibnamefont {Aghassi-Hagmann}}, \bibinfo {author} {\bibfnamefont
  {C.}~\bibnamefont {K\"{u}bel}}, \bibinfo {author} {\bibfnamefont
  {W.}~\bibnamefont {Wulfhekel}}, \bibinfo {author} {\bibfnamefont {I.~M.}\
  \bibnamefont {Pop}},\ and\ \bibinfo {author} {\bibfnamefont {T.}~\bibnamefont
  {Reisinger}},\ }\href {http://arxiv.org/abs/2502.17247} {\bibinfo {title}
  {High quality superconducting tantalum resonators with beta phase defects}}
  (\bibinfo {year} {2025}),\ \bibinfo {note} {arXiv:2502.17247
  [quant-ph]}\BibitemShut {NoStop}%
\bibitem [{\citenamefont {Blanton}\ \emph {et~al.}(1991)\citenamefont
  {Blanton}, \citenamefont {Lelental},\ and\ \citenamefont
  {Barnes}}]{blanton_use_1991}%
  \BibitemOpen
  \bibfield  {author} {\bibinfo {author} {\bibfnamefont {T.}~\bibnamefont
  {Blanton}}, \bibinfo {author} {\bibfnamefont {M.}~\bibnamefont {Lelental}},\
  and\ \bibinfo {author} {\bibfnamefont {C.}~\bibnamefont {Barnes}},\
  }\bibfield  {title} {\bibinfo {title} {The use of {X}-ray diffraction rocking
  curve methodology for assessment of the c-axis orientation in
  {Bi}{Sr}{Ca}{Cu}{O} superconducting thin films},\ }\href
  {https://linkinghub.elsevier.com/retrieve/pii/0921453491915082} {\bibfield
  {journal} {\bibinfo  {journal} {Physica C: Superconductivity and its
  Applications}\ }\textbf {\bibinfo {volume} {184}},\ \bibinfo {pages} {119}
  (\bibinfo {year} {1991})}\BibitemShut {NoStop}%
\bibitem [{\citenamefont {Krinner}\ \emph {et~al.}(2022)\citenamefont
  {Krinner}, \citenamefont {Lacroix}, \citenamefont {Remm}, \citenamefont
  {Di~Paolo}, \citenamefont {Genois}, \citenamefont {Leroux}, \citenamefont
  {Hellings}, \citenamefont {Lazar}, \citenamefont {Swiadek}, \citenamefont
  {Herrmann}, \citenamefont {Norris}, \citenamefont {Andersen}, \citenamefont
  {M\"{u}ller}, \citenamefont {Blais}, \citenamefont {Eichler},\ and\
  \citenamefont {Wallraff}}]{krinner_realizing_2022}%
  \BibitemOpen
  \bibfield  {author} {\bibinfo {author} {\bibfnamefont {S.}~\bibnamefont
  {Krinner}}, \bibinfo {author} {\bibfnamefont {N.}~\bibnamefont {Lacroix}},
  \bibinfo {author} {\bibfnamefont {A.}~\bibnamefont {Remm}}, \bibinfo {author}
  {\bibfnamefont {A.}~\bibnamefont {Di~Paolo}}, \bibinfo {author}
  {\bibfnamefont {E.}~\bibnamefont {Genois}}, \bibinfo {author} {\bibfnamefont
  {C.}~\bibnamefont {Leroux}}, \bibinfo {author} {\bibfnamefont
  {C.}~\bibnamefont {Hellings}}, \bibinfo {author} {\bibfnamefont
  {S.}~\bibnamefont {Lazar}}, \bibinfo {author} {\bibfnamefont
  {F.}~\bibnamefont {Swiadek}}, \bibinfo {author} {\bibfnamefont
  {J.}~\bibnamefont {Herrmann}}, \bibinfo {author} {\bibfnamefont {G.~J.}\
  \bibnamefont {Norris}}, \bibinfo {author} {\bibfnamefont {C.~K.}\
  \bibnamefont {Andersen}}, \bibinfo {author} {\bibfnamefont {M.}~\bibnamefont
  {M\"{u}ller}}, \bibinfo {author} {\bibfnamefont {A.}~\bibnamefont {Blais}},
  \bibinfo {author} {\bibfnamefont {C.}~\bibnamefont {Eichler}},\ and\ \bibinfo
  {author} {\bibfnamefont {A.}~\bibnamefont {Wallraff}},\ }\bibfield  {title}
  {\bibinfo {title} {Realizing repeated quantum error correction in a
  distance-three surface code},\ }\href
  {https://www.nature.com/articles/s41586-022-04566-8} {\bibfield  {journal}
  {\bibinfo  {journal} {Nature}\ }\textbf {\bibinfo {volume} {605}},\ \bibinfo
  {pages} {669} (\bibinfo {year} {2022})}\BibitemShut {NoStop}%
\bibitem [{\citenamefont {Vall\'{e}s-Sanclemente}\ \emph
  {et~al.}(2023)\citenamefont {Vall\'{e}s-Sanclemente}, \citenamefont {Van
  Der~Meer}, \citenamefont {Finkel}, \citenamefont {Muthusubramanian},
  \citenamefont {Beekman}, \citenamefont {Ali}, \citenamefont {Marques},
  \citenamefont {Zachariadis}, \citenamefont {Veen}, \citenamefont {Stavenga},
  \citenamefont {Haider},\ and\ \citenamefont
  {DiCarlo}}]{valles-sanclemente_post-fabrication_2023}%
  \BibitemOpen
  \bibfield  {author} {\bibinfo {author} {\bibfnamefont {S.}~\bibnamefont
  {Vall\'{e}s-Sanclemente}}, \bibinfo {author} {\bibfnamefont {S.~L.~M.}\
  \bibnamefont {Van Der~Meer}}, \bibinfo {author} {\bibfnamefont
  {M.}~\bibnamefont {Finkel}}, \bibinfo {author} {\bibfnamefont
  {N.}~\bibnamefont {Muthusubramanian}}, \bibinfo {author} {\bibfnamefont
  {M.}~\bibnamefont {Beekman}}, \bibinfo {author} {\bibfnamefont
  {H.}~\bibnamefont {Ali}}, \bibinfo {author} {\bibfnamefont {J.~F.}\
  \bibnamefont {Marques}}, \bibinfo {author} {\bibfnamefont {C.}~\bibnamefont
  {Zachariadis}}, \bibinfo {author} {\bibfnamefont {H.~M.}\ \bibnamefont
  {Veen}}, \bibinfo {author} {\bibfnamefont {T.}~\bibnamefont {Stavenga}},
  \bibinfo {author} {\bibfnamefont {N.}~\bibnamefont {Haider}},\ and\ \bibinfo
  {author} {\bibfnamefont {L.}~\bibnamefont {DiCarlo}},\ }\bibfield  {title}
  {\bibinfo {title} {Post-fabrication frequency trimming of coplanar-waveguide
  resonators in circuit {QED} quantum processors},\ }\href
  {https://pubs.aip.org/apl/article/123/3/034004/2903208/Post-fabrication-frequency-trimming-of-coplanar}
  {\bibfield  {journal} {\bibinfo  {journal} {Applied Physics Letters}\
  }\textbf {\bibinfo {volume} {123}},\ \bibinfo {pages} {034004} (\bibinfo
  {year} {2023})}\BibitemShut {NoStop}%
\bibitem [{\citenamefont {Peiniger}\ and\ \citenamefont
  {Piel}(1985)}]{peiniger_superconducting_1985}%
  \BibitemOpen
  \bibfield  {author} {\bibinfo {author} {\bibfnamefont {M.}~\bibnamefont
  {Peiniger}}\ and\ \bibinfo {author} {\bibfnamefont {H.}~\bibnamefont
  {Piel}},\ }\bibfield  {title} {\bibinfo {title} {A {Superconducting} {Nb3Sn}
  {Coated} {Multicell} {Accelerating} {Cavity}},\ }\href
  {https://ieeexplore.ieee.org/document/4334443/} {\bibfield  {journal}
  {\bibinfo  {journal} {IEEE Transactions on Nuclear Science}\ }\textbf
  {\bibinfo {volume} {32}},\ \bibinfo {pages} {3610} (\bibinfo {year}
  {1985})}\BibitemShut {NoStop}%
\end{thebibliography}%

\end{document}